\documentclass[a4paper,11pt]{article}

\usepackage{amsmath}
\usepackage{amssymb}
\usepackage{amsfonts}
\usepackage{verbatim}
\usepackage{graphicx}
\usepackage{epstopdf}
\usepackage{caption}
\usepackage{enumitem}
\usepackage{color,xcolor}
\usepackage[pdfstartview=FitH,
            bookmarksnumbered=true,
            bookmarksopen=true,
            colorlinks=true,
            linkcolor=blue,
            citecolor=blue,
            urlcolor=blue,
            anchorcolor=blue
            ]{hyperref}

\author{Pu Zhang \footnote{Email: puz1@pitt.edu} \ and William J. Parnell \footnote{Email:William.Parnell@manchester.ac.uk} \\
\footnotesize{School of Mathematics, University of Manchester, Oxford Road, Manchester, M13 9PL,UK}}

\title{Band Gap Formation and Tunability in Stretchable Serpentine Interconnects}

\begin{document}

\maketitle

\begin{abstract}
Serpentine interconnects are highly stretchable and frequently used in flexible electronic systems. In this work, we show that the undulating geometry of the serpentine interconnects will generate phononic band gaps to manipulate elastic wave propagation. The interesting effect of `bands-sticking-together' is observed. We further illustrate that the band structures of the serpentine interconnects can be tuned by applying pre-stretch deformation. The discovery offers a way to design stretchable and tunable phononic crystals by using metallic interconnects instead of the conventional design with soft rubbers and unfavorable damping.
\end{abstract}

\section{Introduction}

Phononic crystals are periodic structures/materials exhibiting phononic band gaps that prohibit wave propagation at certain ranges of frequencies \cite{hussein2014dynamics, laude2015phononic, zhang2013broadband}. Conventional phononic crystals are usually composed of rigid materials like metals, plastics, and ceramics so that the band gaps are fixed once the medium is fabricated. In recent years, tunable phononic crystals \cite{barnwell2016inplane, bertoldi2008wave, galich2016elastic, robillard2009tunable, Shmuel2013, ZhangParnell2017} have been of great interest to researchers since the band gaps can be tuned by applying mechanical loads or electrical/magnetic fields \cite{ma2016acoustic}. Currently, most tunable phononic crystals adopt the use of compliant materials like elastomers and gels. However, a critical problem is that these compliant materials exhibit intrinsic damping, which causes undesirable wave dissipation so that high frequency pass bands will usually be eliminated.  This has become an obstacle to the application of tunable phononic crystals since they only work in the very low frequency range ($\sim$ 1 KHz)  \cite{babaee2015harnessing}.

A question that arises therefore is whether it is possible to design tunable phononic crystals by using metals in some fashion to eliminate the material damping. The metallic interconnects in stretchable electronic systems \cite{lu2014flexible, lv2014archimedean, rogers2010materials, zhang2013mechanics} shed some light on this question. These serpentine interconnects are supremely stretchable without inducing irreversible plastic deformation or failure due to their unique deformation feature: large rotation but small strain. Therefore, if the periodic geometry of the serpentine interconnects can form band gaps, they would be very useful in designing stretchable phononic crystals. The interesting work by Trainiti et al.\ \cite{trainiti2015wave} has shown that slightly undulating beams and plates can give rise to phononic band gaps, which implies that the serpentine interconnects may have similar behavior. Researchers have also shown that the resonance frequencies \cite{becker2002natural, pearson1982transfer} and dispersion curves \cite{frikha2011effect} of helical springs can be tuned by applying axial tension/compression, implying the possible tunability of the dynamic behavior for associated types of flexible structures. The main aims of this paper are to explore whether band gaps can be formed by serpentine interconnects and furthermore if these band gaps are tunable by applying a pre-stretch deformation to the stationary state.

The Kirchhoff elastica model \cite{audoly2010elasticity} is used in order to describe the deformation of the serpentine interconnects. Band structures are calculated numerically by solving the incremental dynamic equations that arise when small amplitude deformation is superimposed upon an initially pre-stretch configuration of the serpentine interconnect structure. The tunability of the band gaps under pre-stretch deformation will be discussed. The work provides evidence of the capability to design tunable phononic crystals from serpentine interconnects, which overcomes the material damping obstacle of current designs that employ compliant materials such as elastomers and gels.

\section{Serpentine Interconnects under Stretch}
\begin{figure}[thbp]
	\centering
	\includegraphics [width=9cm] {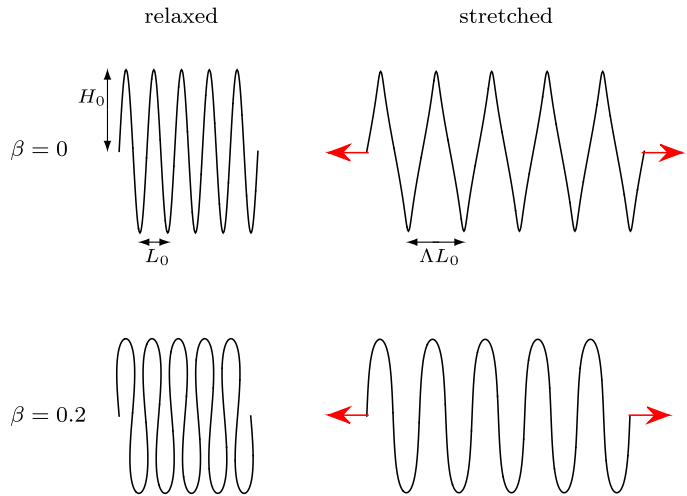}
	\caption{\label{fig:interconnect} Schematic illustration of two forms of stretchable serpentine interconnects. The relaxed configurations are described by Eqs.\ \eqref{x0} and \eqref{y0}.}
\end{figure}

Serpentine interconnects are usually designed as thin undulating elastica, as shown in Fig.\ \ref{fig:interconnect}. These structures exhibit ultra-high stretchability without inducing failure or plastic deformation. As illustrated in Fig.\ \ref{fig:interconnect}, the relaxed configuration is described by parametric curves of the form
\begin{align}
	x_0(s(\xi)) &= L_0 [\xi - \beta \sin(4 \pi \xi)], \label{x0}\\
	y_0(s(\xi)) &= H_0 \sin (2 \pi \xi), \label{y0}
\end{align}
where $x_0$ and $y_0$ are the horizontal and vertical coordinates of the elastica, $H_0$ and $L_0$ are the initial amplitude and period of the undulating elastica, $\beta$ is a parameter indicating the winding degree, $s$ indicates the arc length, and $\xi$ is an independent parameter with $\xi \in [0,1]$ indicating one period. Note that the curve degenerates to a sine function if $\beta = 0$. The relation between the arc length $s$ and coordinate $\xi$ is
\begin{equation} \label{s}
	s(\xi) = \int_{0}^{\xi} \sqrt{x_{0,\zeta}^2 + y_{0,\zeta}^2} \ d \zeta.
\end{equation}
Clearly, the total arc length of one period is obtained as $S = s(1)$ from Eq.\ \eqref{s}. The stretch ratio is $\Lambda$ for the two types of interconnects shown in Fig.\ \ref{fig:interconnect} so the periodic length $L_0$ in the relaxed configuration is stretched to $\Lambda L_0$ in the deformed configuration. We adopt the inextensible Kirchhoff elastica model to describe the deformation and equilibrium of the interconnects.

\subsection{Relaxed Configuration}
\begin{figure}[thbp]
	\centering
	\includegraphics [width=9cm] {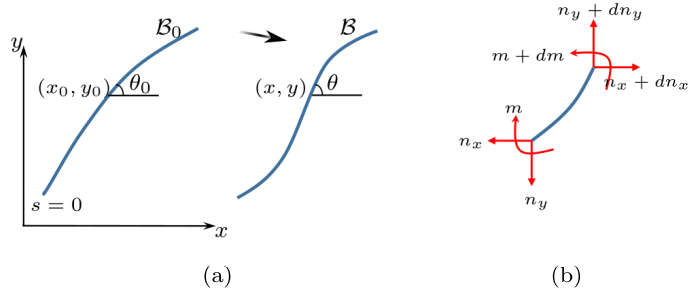}
	\caption{\label{fig:config} Schematic illustration to the relaxed and stretched configurations for the serpentine interconnect.}
\end{figure}

As shown in Fig.\ \ref{fig:config}, the relaxed configuration $ \mathcal{B}_0 $ of the serpentine interconnect is described by the centerline $\mathbf{r}_0(s)=x_0(s) \mathbf{e}_x + y_0(s) \mathbf{e}_y$ such that $s$ is the arclength of the elastica, $\mathbf{e}_x$ and $\mathbf{e}_y$ are the unit bases of the Cartesian coordinate system.
Since the metallic interconnects are usually very thin, the elastica is assumed to be inextensible and shear deformation is negligible. The rotation of an elastica segment is described by $\theta_0$, which is defined as the angle between the centerline and the $\mathbf{e}_x$ direction. The kinematic relation requires that \cite{audoly2010elasticity}
\begin{align}
    x'_0(s) &= \cos \theta_0(s), \label{x0s} \\
    y'_0(s) &= \sin \theta_0(s), \label{y0s}
\end{align}
where $f'(s)=df/ds$. The relaxed configuration $(x_0,y_0)$ is prescribed initially in Eqs. \eqref{x0} and \eqref{y0} and the curvature of the centerline is defined as
\begin{equation} \label{kappa0}
	\kappa_0 (s) = \theta_0'(s).
\end{equation}
Since the parameter $\xi$ is introduced in order to describe the relaxed configuration conveniently, the curvature in Eq. \eqref{kappa0} is further expressed as
\begin{equation} \label{kappa0_new}
	\kappa_0 (s(\xi)) = \frac{x_{0,\xi} y_{0,\xi \xi} - y_{0,\xi} x_{0,\xi \xi}}{(x_{0,\xi}^2 + y_{0,\xi}^2)^{3/2}}.
\end{equation}
The function $\kappa_0(s)$ in Eq. \eqref{kappa0_new} can be solved numerically from Eqs.\ \eqref{s} and \eqref{kappa0_new} together, which will be used later.

\subsection{Stretched Configuration}
As the serpentine interconnect is stretched to a deformed configuration $ \mathcal{B} $ in Fig.\ \ref{fig:config}, the centerline $\mathbf{r}_0$ will be transformed to the current position $\mathbf{r}(s) = x(s) \mathbf{e}_x + y(s) \mathbf{e}_y$. Correspondingly, the rotation angle and curvature also change to the current values as $ \theta(s) $ and $ \kappa(s) $, respectively. Similar to Eqs.\ \eqref{x0s} and \eqref{y0s}, the kinematic relations obey the following form in the stretched configuration,
\begin{align}
	x'(s) &= \cos \theta(s), \label{x} \\
	y'(s) &= \sin \theta(s). \label{y}
\end{align}

The internal forces of the serpentine interconnect include stretching forces $(n_x,n_y)$ and bending moment $ m $ in the deformed configuration, as shown in Fig.\ \ref{fig:config} (b). The dynamic equations are \cite{neukirch2012vibrations}
\begin{align}
    n'_x &= \rho A \ddot{x}, \label{dyn_eq_nx}\\
    n'_y &= \rho A \ddot{y}, \label{dyn_eq_ny}\\
    m' &= n_x \sin \theta - n_y \cos \theta + \rho I \ddot{\theta}, \label{dyn_eq_m}
\end{align}
where $ \dot{f} = \partial f / \partial t $, $t$ denotes time, $ A $ and $ I $ are the area and moment of inertia of the cross section. Finally, the bending moment $m$ is assumed to be proportional to the curvature increment, so that
\begin{equation} \label{cons_eq_m}
    \theta' \equiv \kappa = \kappa_0 + m/(EI),
\end{equation}
where $\kappa $ is the current curvature and $ E $ is the Young's modulus.

The position and internal forces of the stretched configuration can be represented by a state vector $ \boldsymbol{\phi} = (x,y,\theta,n_x,n_y,m) $. Hence, the stretched configuration can be solved by using Eqs.\ \eqref{x} - \eqref{cons_eq_m}, together with the boundary conditions $ \boldsymbol{\phi} (0) $ and $ \boldsymbol{\phi}(S) $ for one period of the serpentine interconnect.

\section{Incremental Dynamic Equation}

We consider small amplitude wave propagation superimposed on a pre-stretch configuration $ \mathcal{B} $, which is called \textit{small on large} in some literature \cite{ogden2007incremental}. Hence the current configuration is a superposition of two terms, as
\begin{equation} \label{phi}
	\boldsymbol{\phi} = \bar{\boldsymbol{\phi}} + \delta \hat{\boldsymbol{\phi}},
\end{equation}
where $ \bar {\boldsymbol \phi} = (\bar{x}, \bar{y}, \bar{\theta}, \bar{n}_x, \bar{n}_y, \bar{m}) $ and $ \hat {\boldsymbol \phi} = (\hat{x}, \hat{y}, \hat{\theta}, \hat{n}_x, \hat{n}_y, \hat{m}) $ represent the equilibrium pre-stretch deformation and the perturbed wave term, respectively.

The pre-stretch deformation $ \bar{\boldsymbol{\phi}} $ can be computed in the following manner. By ignoring the inertia terms, Eqs.\ \eqref{x} - \eqref{cons_eq_m} can be cast into a series of first order ordinary differential equations as $ \bar{\boldsymbol{\phi}}' = \psi (\bar{\boldsymbol{\phi}}) $. The boundary conditions are then specified as
\begin{align}
	s &= 0: \ \bar{x}=\bar{y}=\bar{m}=0, \\
	s &= S: \ \bar{x}=\Lambda L_0, \ \ \bar{n}_y=\bar{m}=0,
\end{align}
where $\Lambda$ is the pre-stretch ratio of the serpentine interconnect, as depicted in Fig.\ \ref{fig:interconnect}. The pre-stretch deformation can then be obtained by solving the equation $ \bar{\boldsymbol{\phi}}' = \psi (\bar{\boldsymbol{\phi}}) $ numerically via the finite difference method. Note that for the relaxed configuration chosen in Eqs.\ \eqref{x0} and \eqref{y0}, the internal forces $\bar{n}_x = \textrm{const.}$ and $\bar{n}_y = 0$.

After substituting the assumed perturbed solution \eqref{phi} into the six equations Eqs.\ \eqref{x} - \eqref{cons_eq_m}, we derive the governing equations for the wave propagation problem (see Appendix \ref{appendix_A}), as
\begin{equation}\label{wave_eq}
	\hat{\boldsymbol \phi}' = \mathbf{K} \hat{\boldsymbol \phi} + \mathbf{M} \ddot{\hat{\boldsymbol{\phi}}},
\end{equation}
where $\mathbf{K}$ and $\mathbf{M}$ are 6-by-6 matrices with nonzero components given by
\begin{gather*}
	K_{13} = -\sin \bar{\theta},\  K_{23} = \cos \bar{\theta},\  K_{36}=1/(EI),\\
	K_{63} = \bar{n}_x \cos \bar{\theta} + \bar{n}_y \sin \bar{\theta}, \ K_{64} = \sin \bar{\theta},\  K_{65} = -\cos \bar{\theta}, \\
	M_{41} = M_{52} = \rho A, \ M_{63} = \rho I.
\end{gather*}

For the serpentine interconnects, the Bloch wave solution to Eq.\ \eqref{wave_eq} is
\begin{equation} \label{Bloch_sol}
	\hat{\boldsymbol{\phi}} = \boldsymbol{\Phi} e^{i(ks-\omega t)},
\end{equation}
where $\boldsymbol{\Phi} = (X,Y,i \Theta, i N_x, i N_y, M)$ is a periodic complex function of $s$, $k$ is the wave number, and $\omega$ the eigenfrequency. After substituting Eq.\ \eqref{Bloch_sol} into \eqref{wave_eq} we find that
\begin{equation} \label{Bloch_mode}
	\boldsymbol{\Phi}' =\mathbf{A} \boldsymbol{\Phi},
\end{equation}
where $\mathbf{A} = \mathbf{K} - ik \mathbf{I} - \omega^2 \mathbf{M} $ and $\mathbf I$ is the 6-by-6 identity matrix. The transfer matrix method \cite{laude2015phononic} can be used to solve Eq.\ \eqref{Bloch_mode}. In order to implement this method, the arc length $S$ in one period is divided into $P$ equal intervals \cite{becker2002natural, pearson1982transfer} as $\Delta s = S/P$ with $P$ large enough (e.g.\ $> 200$). Thereby, following the Cayley-Hamilton theorem \cite{bellman1997introduction}, the solution of Eq.\ \eqref{Bloch_mode} for $s=p \Delta s \ (p=1,2,\cdots,P)$ is
\begin{equation} \label{phi_ps}
	\boldsymbol{\Phi}(p  \Delta s) = \prod_{j=1}^{p} \exp(\mathbf{A}_j \Delta s) \boldsymbol{\Phi}(0) = \mathbf{T}_p \boldsymbol{\Phi}(0),
\end{equation}
where $ \mathbf{A}_j = [\mathbf{A}(j \Delta s) + \mathbf{A}((j-1) \Delta s) ]/2 $ and $ \mathbf{T}_p = \mathbf{T}(p \Delta s) $ called the transfer matrix. Finally, the state vector for the end of one period $ s=S $ can be obtained from Eq.\ \eqref{phi_ps}, as
\begin{equation} \label{phi_S}
	\boldsymbol{\Phi}(S) = \mathbf{T}_P \boldsymbol{\Phi}(0).
\end{equation}
Therefore, with Eq.\ \eqref{phi_S}, the Bloch boundary condition $ \boldsymbol{\Phi}(0) = \boldsymbol{\Phi}(S) $ finally leads to the following eigenvalue problem
\begin{equation} \label{det}
	\det (\mathbf{T}_P - \mathbf{I}) = 0.
\end{equation}
The eigenfrequency $\omega(k)$ can be solved numerically from Eq.\ \eqref{det} once a wave number $k \ (0 \leqslant k \leqslant \pi)$ is specified. The real eigenfrequencies $\omega(k)$ represent propagating waves while complex eigenfrequencies indicate evanescent waves. The band structure is usually plotted for the real eigenfrequencies. The Bloch modes can be calculated from Eq.\ \eqref{phi_ps} by replacing $\boldsymbol{\Phi}(0)$ with the eigenvectors of Eq.\ \eqref{det}. Normally either the real or imaginary part of the complex Bloch mode can be used to depict the Bloch wave modes.

\section{Numerical Study}

The serpentine interconnects are highly stretchable due to their undulating geometry. Wave propagation behavior is expected to be influenced by the geometry as well as the pre-stretch ratio. Some of the key questions to be answered include,
\begin{itemize}
	\item Will the undulating geometry of the serpentine interconnect lead to band gap formation, and if yes, how would the geometric parameters affect the the band gap structure?
	\item How will the pre-stretch deformation affect the band gaps?
	\item Which is the dominant factor for the band gap tunability, the undulating geometry or internal force?
\end{itemize}
In order to address these questions, we conduct numerical simulations for the band structure to explore band gap formation and tunability of the serpentine interconnects. Two dimensionless variables are introduced for the wave number $k$ and eigenfrequency $\omega$, as
\begin{equation}
	\bar{k} = kS, \quad \bar{\omega} = \omega S \sqrt{\frac{\rho A}{EI}}.
\end{equation}
In addition, the rotary inertia term $M_{63}$ is also ignored in Eq.\ \eqref{wave_eq} since its effect is negligible for thin rods.

\begin{figure}[thbp]
	\centering
	\includegraphics [width=8cm] {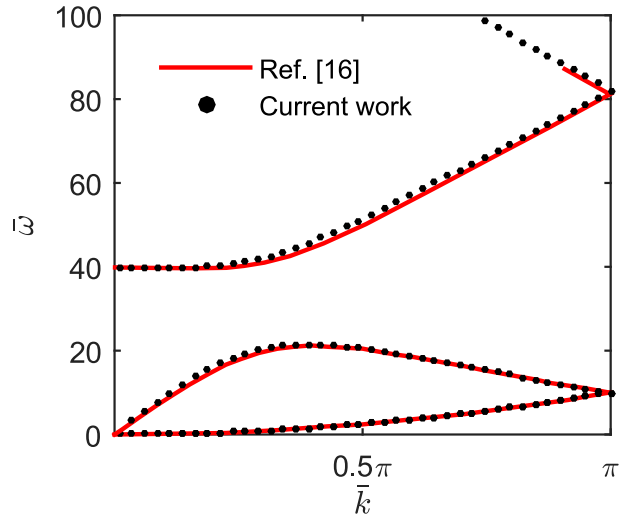}
	\caption{\label{fig:compare} Comparison of the dispersion curves obtained from the current elastica model and the undulated beam model \cite{trainiti2015wave}. The geometric parameters are chosen as $H_0/L_0 = 0.05, \beta=0$ and the thickness of the beam is $0.07L_0$. }
\end{figure}
The current model is compared with the undulated beam model proposed in Ref. \cite{trainiti2015wave} to verify the results. Figure \ref{fig:compare} illustrates the dispersion curves of a slightly undulated beam ($ H_0/L_0=0.05, \beta=0 $) obtained from these two models. It can be seen that these two models predict almost identical results in this case. In fact, the elastica model degenerates to the undulated beam model when the amplitude $H_0$ is much smaller than $L_0$. Note that even though the undulated beam model \cite{trainiti2015wave} is relatively easy to use, it is only applicable when the beam is slightly undulated, otherwise significant error will be induced. Since serpentine interconnects usually exhibit severely undulated geometry incorporated by the presence of the additional winding parameter, the elastica model must be used to determine its deformation \cite{audoly2010elasticity,zhang2013mechanics} and its dynamic behaviour \cite{neukirch2012vibrations}.

\begin{figure}[thbp]
	\centering
	\includegraphics [width=16cm] {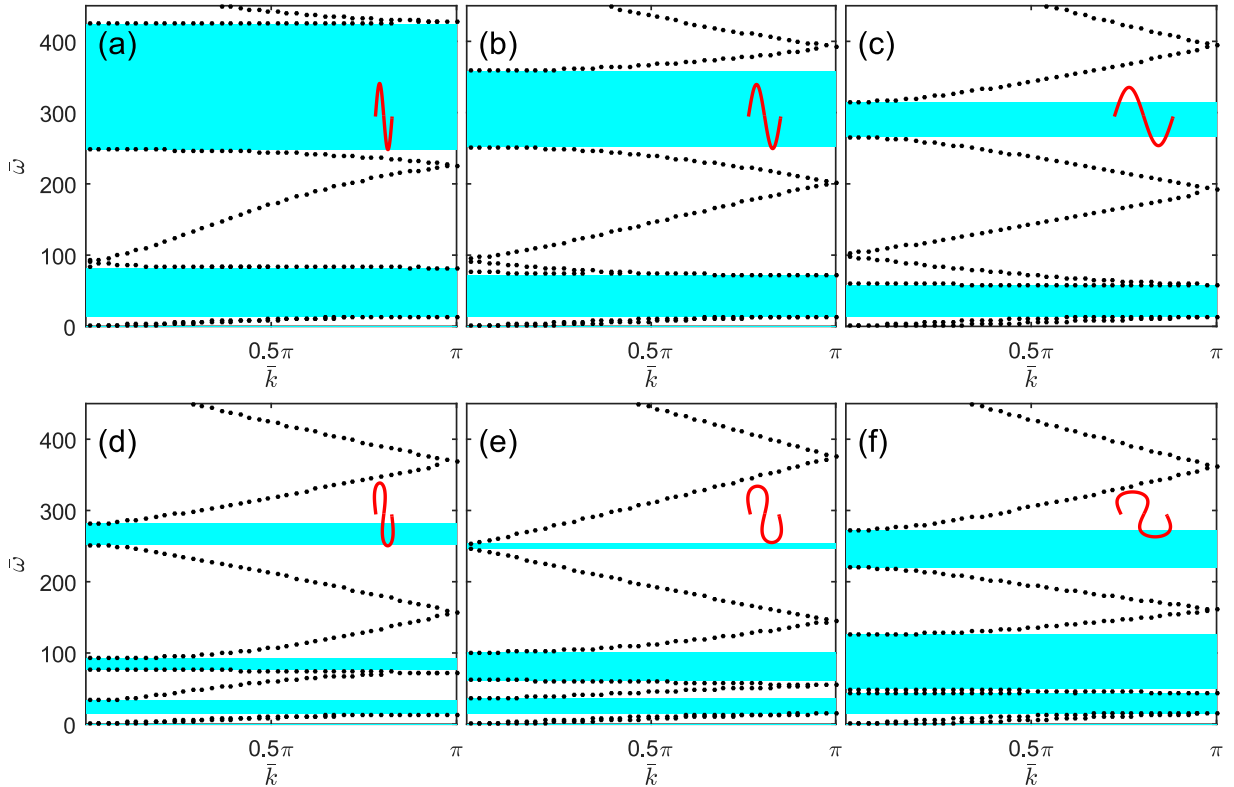}
	\caption{\label{fig:relaxed} Phononic band structures of serpentine interconnects in the relaxed state. The dispersion curves are indicated by dotted curves, while the band gaps are shaded in cyan (online) or grey (print). (a) $ H_0/L_0=2, \ \beta=0 $, (b) $ H_0/L_0=1, \ \beta=0 $, (c) $ H_0/L_0=0.5, \ \beta=0 $, (d) $ H_0/L_0=2, \ \beta=0.2 $, (e) $ H_0/L_0=1, \ \beta=0.2 $, (f) $ H_0/L_0=0.5, \ \beta=0.2 $.  }
\end{figure}

The dispersion relations of serpentine interconnects without pre-stretch are shown in Fig.\ \ref{fig:relaxed}. We consider two types of interconnects with different winding parameter $\beta$. In addition, the aspect ratio $H_0/L_0$ is also varied to investigate its effect on band gap formation. Fig.\ \ref{fig:relaxed} (a)-(c) illustrate the band structures of sinusoidal interconnects ($\beta=0$) with aspect ratios $H_0/L_0 =$ 2, 1, and 0.5, respectively.
It is observed from Fig.\ \ref{fig:relaxed} (a)-(c) that the undulating geometry of the interconnects usually generates multiple band gaps, which are useful for wave filtering and control purposes. This band gap formation mechanism is different to that in conventional two-phase phononic crystals where non-uniform material distribution is employed \cite{hussein2014dynamics, laude2015phononic, zhang2013broadband}. Instead, serpentine interconnects generate band gaps from the non-uniform geometry, similar to sinusoidal beams \cite{trainiti2015wave, maurin2014wave} and undulated lattice structures \cite{trainiti2016wave, chen2017lattice}. According to Fig.\ \ref{fig:relaxed} (a)-(c), associated with the winding parameter $\beta=0$, an increase in the aspect ratio $H_0/L_0$ will widen the first and second band gaps significantly. In the extreme scenario when $H_0/L_0= 0$, i.e.\ a straight elastica, there will be no band gaps.  Figure\ \ref{fig:relaxed} (d)-(e) show the band structures of serpentine interconnects with $\beta=0.2$ for different aspect ratios $H_0/L_0$. Interestingly, for this winding parameter, these band structures have different characteristics as $H_0/L_0$ is increased, compared with those in Fig.\ \ref{fig:relaxed} (a)-(c). In particular as is illustrated in Fig.\ \ref{fig:relaxed} (d)-(e), the opposite effect is noted in the first and second band gaps in that their width now \textit{decreases} as $H_0/L_0$ increases. The advantage of increasing the winding parameter $\beta$ is to enhance the stretchability of the serpentine interconnects, thinking ahead to its potential as a tunable structure. The numerical analysis in Fig.\ \ref{fig:relaxed} answers the first question posed at the beginning of this section. The non-uniform geometry of the serpentine interconnects will induce band gaps naturally without the need to introduce any material inhomogeneity. Additionally, the band gaps can be modified by changing geometric parameters such as the aspect ratio $H_0/L_0$ and the winding parameter $\beta$ of the serpentine interconnects.

A remarkable feature of the band structures of serpentine interconnects is the so-called `bands-sticking-together' effect \cite{dresselhaus2007group}. It can be observed from Fig.\ \ref{fig:relaxed} that all of the bands occur as pairs which merge at $\bar{k}=\pi$, indicating two-fold degeneracies of the Bloch modes at the edge of the first Brillouin zone. Note in particular that the first two bands are almost indistinguishable in Fig.\ \ref{fig:relaxed}. Therefore band gaps can only exist between two bands in neighbour pairs, not between two bands of the same pair. This phenomenon is induced by the intrinsic glide symmetry of the serpentine interconnects. In other words, a reflection operation with respect to the horizontal axis followed by a horizontal translation $L_0/2$ will not change the structure in Fig.\ \ref{fig:interconnect}. In fact the serpentine interconnects belong to the nonsymmorphic group $F_{2mg}$ in the seven Frieze groups \cite{mock2010space}. The `bands-sticking-together' phenomenon was also observed for photonic crystal waveguides \cite{mock2010space} with the same space group. The pre-stretch deformation will not change the underlying symmetry of the serpentine interconnects in this work.

\begin{figure}[thbp]
	\centering
	\includegraphics [width=12cm] {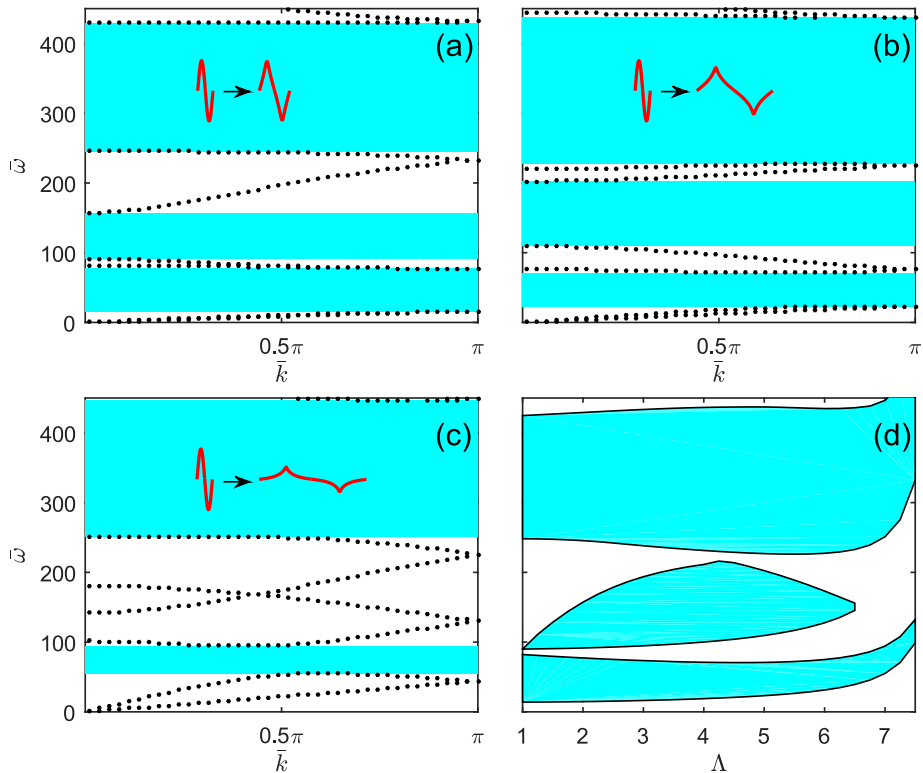}
	\caption{\label{fig:stretch0} Phononic band structures of pre-stretched serpentine interconnects with $\beta=0$. (a) $\Lambda = 2$, (b) $\Lambda = 5$, (c) $\Lambda = 7$, (d) Band gaps for different stretch ratio $\Lambda$. }
\end{figure}

In order to address the second question, we study the effect of pre-stretch deformation on the band structure characteristics and band gap tunability of the serpentine interconnects. The analysis shall focus on interconnects with a high aspect ratio $H_0/L_0=2$. Figure \ \ref{fig:stretch0} shows the band structures of sinusoidal serpentine interconnects ($\beta=0$) under pre-stretch deformation. The variation in band gap width is illustrated in Fig.\ \ref{fig:stretch0} (d) for a wide range of stretch ratios $\Lambda$ up to approximately $7.5$. Several band gap structures for specific $\Lambda$ are shown in Fig. \ref{fig:relaxed} (a) and Figs.\ \ref{fig:stretch0} (a)-(c). It is observed from Fig.\ \ref{fig:stretch0} (d) that for this parameter set, there usually exists three band gaps at frequencies $\bar{\omega}<450$ for most of the stretch ratios considered here. The first band gap shrinks when $\Lambda$ increases while the third band gap widens. The central frequency of the first band gap decreases slightly then increases dramatically as $\Lambda$ becomes larger, which is induced by the separation of the first (and second) pair of bands, and is seen by comparing Fig.\ \ref{fig:stretch0} (a) with Fig.\ \ref{fig:stretch0} (c). Fig.\ \ref{fig:stretch0} (d) also indicates that the second band gap disappears when $\Lambda>6.5$, an extremely stretched situation. For instance, there is no band gap between the second and third pairs of bands in Fig.\ \ref{fig:stretch0}(c). Instead, there exists a crossing between the fourth and fifth bands at $\bar{k}=0.45 \pi$, which therefore prevents the existence of a band gap.

\begin{figure}[thbp]
	\centering
	\includegraphics [width=12cm] {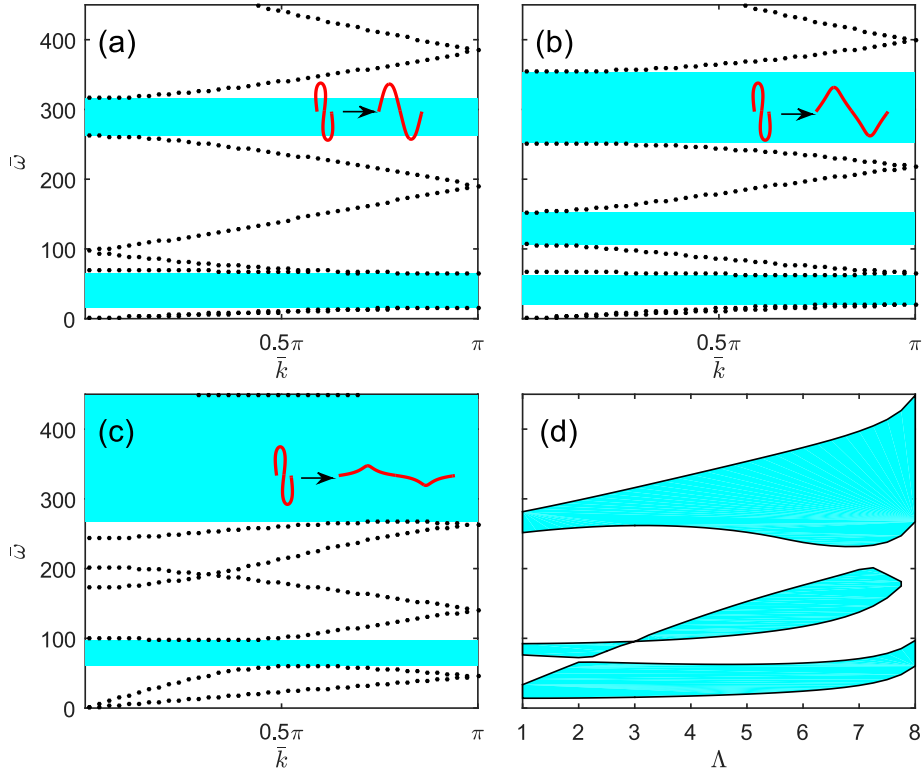}
	\caption{\label{fig:stretch0p2} Phononic band structures of pre-stretched serpentine interconnects with $\beta=0.2$. (a) $\Lambda = 3$, (b) $\Lambda = 5$, (c) $\Lambda = 8$, (d) Band gaps for different stretch ratio $\Lambda$. }
\end{figure}

Figure\ \ref{fig:stretch0p2} depicts the band structures of serpentine interconnects when $\beta=0.2$ with band gaps shown in Fig.\ \ref{fig:stretch0p2} (d) for stretch ratio $\Lambda$ up to $8$. Overall, the effect of pre-stretch deformation on the band structure for this case is more complicated than that of the sinusoidal interconnects. It can be seen from Fig.\ \ref{fig:stretch0p2} (d) that the second band gap disappears when the stretch ratio $\Lambda=3$ (see Fig. \ref{fig:stretch0p2} (a)) and subsequently widens, indicating complicated interactions between the fourth and fifth bands in the band structures. Similar to Fig.\ \ref{fig:stretch0} (d), the second band gap in Fig.\ \ref{fig:stretch0p2} (d) also disappears when the stretch ratio is large enough, due to the band crossing effect, e.g.\ the crossing between the fourth and fifth bands at $\bar k = 0.32 \pi$ which is visible in Fig.\ \ref{fig:stretch0p2} (c). 

Figures \ref{fig:stretch0} and \ \ref{fig:stretch0p2} demonstrate that the band gaps in serpentine interconnects are highly tunable via the application of pre-stretch deformation. Hence, it is possible to design highly stretchable, yet tunable, phononic crystals by using such serpentine interconnects. However, in an extreme scenario when the serpentine interconnects are stretched to straight lines, all band gaps will disappear, although we note that this would be a rare situation in practice. In this case, the serpentine interconnect will behave like straight beams under constant axial tension.

\begin{figure}[thbp]
	\centering
	\includegraphics [width=12cm] {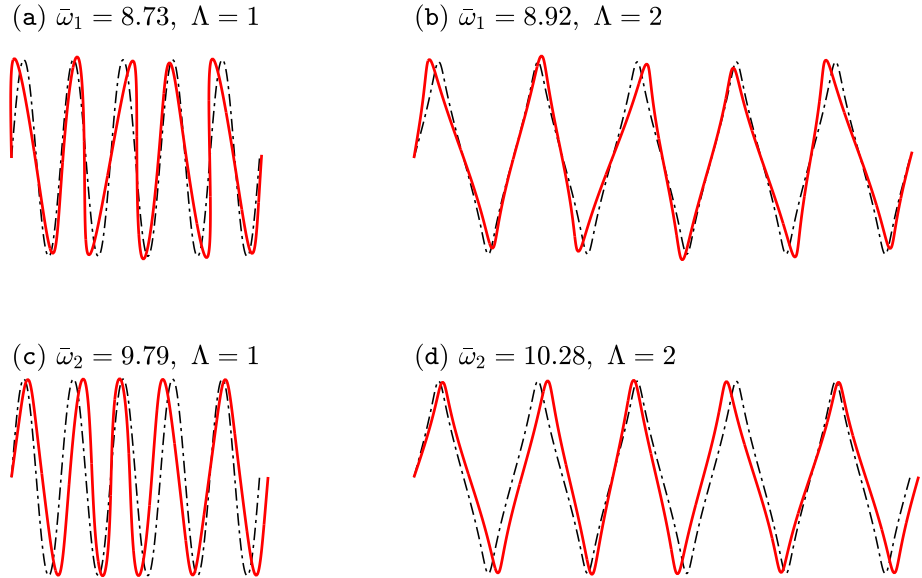}
	\caption{\label{fig:mode} Bloch wave modes for relaxed ((a) and (c)) and pre-stretched ((b) and (d)) serpentine interconnects. The stationary configurations are indicated by dash-dot curves while the Bloch wave modes are represented by solid curves. The wave number is chosen as $\bar{k} = \pi/2$ and the eigenfrequencies $\bar \omega_1$ and $\bar \omega_2$ indicate the first and second bands, respectively. The geometric parameters are chosen as $H_0/L_0=2$ and $\beta=0$ for all cases. }
\end{figure}

For the purposes of illustration, the Bloch wave modes of the relaxed and stretched serpentine interconnects are shown in Fig.\ \ref{fig:mode}. The Bloch mode is obtained by superimposing the real part of the solution $\hat{\boldsymbol{\phi}}$ in Eq.\ \eqref{Bloch_sol} onto the pre-stretch deformation $\bar{\boldsymbol{\phi}}$. Note that the time-dependent term $e^{-i \omega t}$ is omitted in Eq.\ \eqref{Bloch_sol}. In Fig.\ \ref{fig:mode}, we pick up a special scenario with $\bar{k}=\pi/2$ for the first and second bands $\bar{\omega}_1$ and $\bar{\omega_2}$.
The Bloch wave modes in Fig.\ \ref{fig:mode} (a) and (c) correspond to Fig.\ \ref{fig:relaxed} (a), while the two modes in Fig.\ \ref{fig:mode} (b) and (d) correspond to Fig.\ \ref{fig:stretch0}(a). It is found that the first band $\bar \omega_1$ is a flexural mode while the second band $\bar \omega_2$ is longitudinal, which are shown clearly in Fig.\ \ref{fig:mode}.  The pre-stretch deformation does not significantly modify the wave mode characteristics of the serpentine interconnects, while the eigenfrequencies will normally be tuned by the pre-stretch to some extent. Higher order Bloch modes will usually exhibit local bending deformation of the interconnects. In addition, we have found that the serpentine interconnects with winding parameter $\beta=0.2$ show similar Bloch mode characteristics, which will not be discussed further in detail.

\begin{figure}[thbp]
	\centering
	\includegraphics [width=12cm] {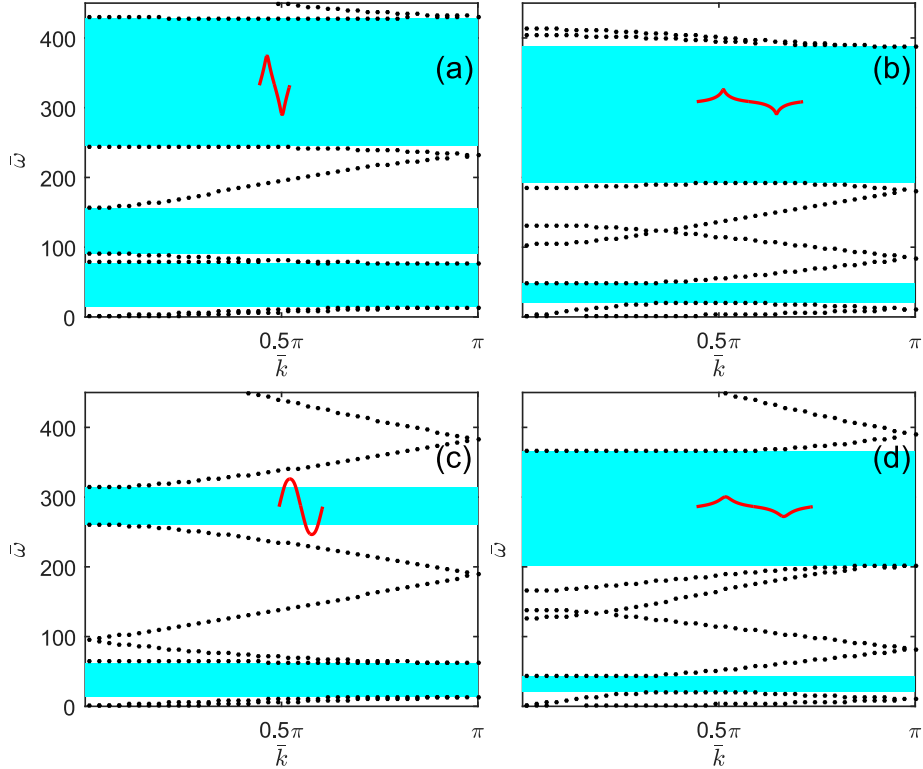}
	\caption{\label{fig:noforce} Phononic band structures of pre-stretched serpentine interconnects when internal forces are neglected. The aspect ratio is $ H_0/L_0=2 $ for all cases. (a) $ \beta=0, \ \Lambda=2 $; (b) $  \beta=0, \ \Lambda=7  $; (c) $\beta=0.2, \ \Lambda=3 $; (d) $ \beta=0.2, \ \Lambda=8  $.}
\end{figure}

The wave dispersion relations of pre-stretched serpentine interconnects are mainly influenced by two factors: the geometry change and internal forces. It would be interesting to quantify these two effects and identify which, if any, are dominant. This is done here by setting $ K_{63}=0 $ in Eq.\ \eqref{wave_eq} so that the effect of internal forces $\bar{n}_x$ and $\bar{n}_y$ is therefore neglected for the subsequent band structure calculation. Fig.\ \ref{fig:noforce} illustrates the band structures of the serpentine interconnects in this case, i.e.\ without the consideration of internal forces. Specifically, Fig.\ \ref{fig:noforce} (a)-(d) correspond to the cases in Fig.\ \ref{fig:stretch0} (a),  \ref{fig:stretch0} (c), \ref{fig:stretch0p2} (a), \ref{fig:stretch0p2} (c), respectively, by considering the geometry change only. By comparing with  Fig.\ \ref{fig:stretch0} (a) and Fig.\ \ref{fig:stretch0p2} (a), it is found from Fig. \ \ref{fig:noforce} (a) and (c) that the band structures are only slightly affected by the internal forces, in the scenario when the stretch ratio $\Lambda$ is not large. In contrast, once the stretch ratio $\Lambda$ is large enough, e.g. in Fig.\ \ref{fig:noforce} (b) and (d), the internal forces play a significant role on the band structures with most bands shifting upward to higher frequencies. Our simulation indicates that the band structures of serpentine interconnects are significantly affected by the internal forces only when the stretch ratio $\Lambda>6$. As a result, the central frequencies of band gaps in Fig. \ref{fig:stretch0} (d) and \ref{fig:stretch0p2} (d) increase dramatically when $\Lambda>6$. Hence, we conclude that the geometry change is the dominant effect on band structure modification when the stretch ratio $\Lambda$ is below the intermediate level. In the case extreme stretch however, both the geometry change and internal forces play significant roles on the band structure. This answers the third question raised above.

\section{Conclusions}
In this work, we have explored the possibility of using serpentine interconnects as stretchable phononic materials to control elastic waves and studied the influence of pre-stretch deformation on the tunability of band gaps. The serpentine interconnects are modeled as inextensible Kirchhoff elastica. Wave propagation is governed by incremental dynamic equations superimposed on a pre-stretch state. The band structures are obtained numerically for both the relaxed and pre-stretched serpentine interconnects. Numerical results indicate that band structures of serpentine interconnects exhibit the `bands-sticking-together' effect due to the intrinsic glide symmetry. All bands occur as pairs and degenerate at the edge of the first Brillouin zone. The band gaps are formed between bands of neighbor pairs and tuned by their strong interactions. In addition, the pre-stretch deformation can tune the band gaps significantly, providing opportunities for wave filtering and control. There are two main factors that affect the band gaps of the pre-stretched serpentine interconnects: geometry change and internal forces. Our results indicate that the geometry change is dominant when the stretch ratio is lower than 5 while both effects play significant roles for extremely stretched scenarios.

The serpentine interconnect therefore offers a way to design stretchable and tunable phononic media with low damping, which is the key advantage over the prevalent design of using compliant materials such as elastomers and gels. Future research could be directed towards experimental studies as well as two- and three-dimensional designs of such phononic materials employing serpentine interconnects. In addition, it would also be very interesting to explore the effects of different geometric design \cite{zhang2013mechanics,lv2014archimedean} on the band gaps and tunability of serpentine interconnects.

\subsection*{Acknowledgments}
The authors are grateful to the Engineering and Physical Sciences Research Council (EPSRC) for financial support via grant no. EP/L018039/1.

\appendix
\numberwithin{equation}{section}
\section{Derivation for Incremental Dynamic Equations} \label{appendix_A}
The derivation of the incremental dynamic equation \eqref{wave_eq} is introduced in this section. After substituting the perturbed solution \eqref{phi} into Eq. \eqref{x} and performing a Taylor expansion with respect to $\delta$, one obtains
\begin{equation} \label{App_x}
\begin{split}
   \bar{x}'+\delta \hat{x}' &= \cos(\bar{\theta}+\delta \hat{\theta}) \\
                            &= \cos \bar{\theta} - \sin \bar{\theta} \ \delta \hat{\theta} + O(\delta^2).
\end{split}
\end{equation}
Note that $\bar{x}'=\cos \bar{\theta}$ is satisfied automatically for the pre-stretch configuration in Eq.\ \eqref{App_x}. Hence, the incremental dynamic equation is obtained by collecting the terms at $O(\delta)$, i.e.\
\begin{equation}
	\hat{x}' = -\sin \bar{\theta} \ \hat{\theta}.
\end{equation}
The incremental dynamic equations corresponding to Eqs. \eqref{y}-\eqref{dyn_eq_ny} and \eqref{cons_eq_m} are derived similarly. Finally, the substitution of \eqref{phi} into \eqref{dyn_eq_m} results in
\begin{equation} \label{app_m}
	\begin{split}
	\bar{m}'+\delta \hat{m}' &= (\bar{n}_x + \delta \hat{n}_x) \sin(\bar{\theta}+\delta \hat{\theta}) - (\bar{n}_y + \delta \hat{n}_y) \cos(\bar{\theta}+\delta \hat{\theta}) + \rho I (\ddot{\bar \theta} + \delta \ddot{\hat \theta}) \\
	&\approx (\bar{n}_x + \delta \hat{n}_x) (\sin \bar{\theta}+\cos \bar{\theta} \delta \hat{\theta}) - (\bar{n}_y + \delta \hat{n}_y) (\cos \bar{\theta}-\sin \bar{\theta} \delta \hat{\theta}) + \rho I (\ddot{\bar \theta} + \delta \ddot{\hat \theta}) \\
	&\approx \bar{n}_x \sin \bar{\theta} - \bar{n}_y \cos \bar{\theta} + \rho I \ddot{\bar{\theta}} + \sin \bar{\theta} \delta \hat{n}_x -\cos \bar{\theta} \delta \hat{n}_y +(\bar{n}_x \cos \bar{\theta}+\bar{n}_y \sin \bar{\theta})\delta \hat{\theta} + \rho I \delta \ddot{\hat{\theta}}.
	\end{split}
\end{equation}
By collecting the terms of $O(\delta)$ in Eq. \eqref{app_m}, we obtain the incremental dynamic equation as
\begin{equation}
	\hat{m}' = \sin \bar{\theta} \hat{n}_x -\cos \bar{\theta} \hat{n}_y +(\bar{n}_x \cos \bar{\theta}+\bar{n}_y \sin \bar{\theta}) \hat{\theta} + \rho I \ddot{\hat{\theta}}.
\end{equation}
Note that the inertia term $\rho I \ddot{\bar \theta}= 0$ for the stationary pre-stretch state. Finally, the incremental dynamic equations can be expressed in a matrix form as shown in Eq.\ \eqref{wave_eq}.

{\small
	
\bibliographystyle{unsrt}

}

\end{document}